\documentclass[a4paper,10pt]{revtex4}

\usepackage{graphicx} 
\newcommand{\be}{\begin{equation}}
\newcommand{\ee}{\end{equation}} 
\catcode `\@=11 %
\font\fivegr@@k=cmmi5 %
\font\sevengr@@k=cmmi7 %
\newdimen\d@pthcorr \d@pthcorr=1pt
\newdimen\prop@rshift
\newdimen\taild@pth
\def\@@lap#1{\vbox to 0pt{\hbox{\raise\d@pthcorr\llap{#1}}\vss}}%
\def\k@@pk@rns#1#2{%
#1{\setbox0=\hbox{#2}\kern-\wd0\copy0\kern-\wd0}#1}%
\def\gr@@kchar{\noindent\char'023\kern\prop@rshift}%
\def\s@tprop@rshift#1{\prop@rshift=#1em}
\def\stor@curr@ntfont{\relax}
\def\s@tprop@rfont#1{
\ifnum \the\lccode`#1=`#1 
\fivegr@@k\else%
\ifnum \the\uccode`#1=`#1 
\sevengr@@k%
\fi\fi%
\taild@pth=1ex \advance\taild@pth by -\d@pthcorr}
\def\tail@d#1#2{%
#1
\stor@curr@ntfont%
\s@tprop@rshift{#2}%
{\s@tprop@rfont{#1}\setbox0=\@@lap{\gr@@kchar}%
\dp0=\taild@pth \box0}}%
\def\Polisha{\tail@d{a}{0.0}}%
\def\PolishA{\k@@pk@rns{A}{\tail@d{A}{0.05}}}%
\def\Polishe{\tail@d{e}{0.05}}%
\def\PolishE{\tail@d{E}{0.05}}%
\def\supr@ss#1#2{
\leavevmode \rlap{\kern#1\char'40}#2}
\def\Polishl{\leavevmode \/\supr@ss{0em}l}
\def\PolishL{\supr@ss{0.1em}L}%
\newdimen\polop@nquotdim%
\def\polopenquot{%
\setbox0=\hbox{''}\polop@nquotdim=\ht0%
\setbox0=\hbox{,}\advance\polop@nquotdim by -\ht0%
\setbox1=\hbox{\lower\polop@nquotdim\hbox{''}}%
\ht1=\ht0 \dp1=\dp0%
\box1}%
\def\@@symbol{\vrule height 1ex width 0.2em depth -0.9ex%
\llap{\char'27 \kern -0.28em }\hskip 0.1em}%
\def\numbersymbol{N\kern0.05em\@@symbol}%
\catcode `\@=12%
\let\previoustwobackslashes=\\%
\def\\#1{%
\if      a#1\Polisha%
\else\if c#1\'c%
\else\if e#1\Polishe%
\else\if l#1\Polishl%
\else\if n#1\'n%
\else\if o#1\'o%
\else\if s#1\'s%
\else\if x#1\'z%
\else\if z#1\.z%
\else\if A#1\PolishA%
\else\if C#1\'C%
\else\if E#1\PolishE%
\else\if L#1\PolishL%
\else\if N#1\'N%
\else\if O#1\'O%
\else\if S#1\'S%
\else\if X#1\'Z%
\else\if Z#1\.Z%
\else\if @#1\numbersymbol%
\else\if ,#1\polopenquot%
\else%
\ifhmode %
        \hskip2pt \vrule height1.5ex width0.5em depth0.2ex \hskip2pt %
\else %
        \ifvmode %
                \vskip2pt \hrule height1.5ex width0.5em depth0.2ex \vskip2pt %
        \fi %
\fi %
\message{\string\\#1 - NIE MA TAKIEGO ZNAKU}%
\fi\fi\fi\fi\fi\fi\fi\fi\fi\fi%
\fi\fi\fi\fi\fi\fi\fi\fi\fi\fi}%
\def\normalslash{/}
\begingroup
\catcode`\/=\active
\global\def\switchtoslash{%
\catcode`\/=\active%
\def/##1{%
\ifx     /##1\normalslash
\else\if a##1\Polisha%
\else\if c##1\'c%
\else\if e##1\Polishe%
\else\if l##1\Polishl%
\else\if n##1\'n%
\else\if o##1\'o%
\else\if s##1\'s%
\else\if x##1\'z%
\else\if z##1\.z%
\else\if A##1\PolishA%
\else\if C##1\'C%
\else\if E##1\PolishE%
\else\if L##1\PolishL%
\else\if N##1\'N%
\else\if O##1\'O%
\else\if S##1\'S%
\else\if X##1\'Z%
\else\if Z##1\.Z%
\else\if @##1\numbersymbol%
\else\if ,##1\polopenquot%
\else%
\ifhmode %
        \hskip2pt \vrule height1.5ex width0.5em depth0.2ex \hskip2pt %
\else %
        \ifvmode %
                \vskip2pt \hrule height1.5ex width0.5em depth0.2ex \vskip2pt %
        \fi %
\fi %
\message{\string/##1 - NIE MA TAKIEGO ZNAKU}%
\fi\fi\fi\fi\fi\fi\fi\fi\fi\fi\fi%
\fi\fi\fi\fi\fi\fi\fi\fi\fi\fi%
}%
\let\\=\previoustwobackslashes}%
\endgroup
\switchtoslash

\begin{document}

\title{ The durations of recession and prosperity: \\ does their distribution
follow a power or an exponential law?} 
\author{ Marcel Ausloos$^{1,}$ 
Janusz Mi\'skiewicz$^{1,2, \dagger}$, 
Mich$\grave e$le Sanglier$^{3, \dagger\dagger}$ }

\email[]{marcel.ausloos@ulg.ac.be}

\email[$^{\dagger}$]{jamis@ozi.ar.wroc.pl}

\email[$^{\dagger\dagger}$]{msanglie@ulb.ac.be}

\affiliation{ $^{1}$ SUPRATECS, Institute of  Physics, B5, University of
Li$\grave e$ge, B-4000 Li$\grave e$ge, Euroland \\ $^{2}$ University of
Agriculture, Department of Physics and Biophysics, ul. Norwida 25, 50-375
Wroc\l{}aw, Poland \\ $^{3}$ Facult\'e des Sciences, Cellule de
Mod\'elisation de la Complexit\'e en Sciences Sociales ULB et Instituts Solvay,
Universit\'e Libre de Bruxelles, Campus Plaine CP 231, B-1050 
Bruxelles, Belgium}

\begin{abstract} 
Following findings by Ormerod and Mounfield 
\cite{ormer} Wright
\cite{wright} rises the problem whether a power \cite{ormer} or an 
exponential law
\cite{wright} describes the distribution of occurrences of economic recession
periods. In order to clarify the controversy a different set of GDP data is
hereby examined. The conclusion about a power law distribution of recession
periods seems better though the matter is not entirely settled. The case of
prosperity duration is also studied and is found to follow a power 
law. Universal
but also non universal features between recession and prosperity cases are
emphasized. Considering that the economy is basically a bistable
(recession//prosperity) system we may derive a characteristic (de)stabilisation
time. 
\end{abstract}

\maketitle

\section{Introduction}

Ormerod and Mounfield \cite{ormer} have analysed data from 17 
capitalist economies
between 1870 and 1994 and concluded that the {\it number} of duration of
recessions is consistent with a power law. Wright \cite{wright} claims that the
data rather follows an exponential law. However the problem, i.e. which law is
really governing the occurrence distribution of recessions was not solved by
Wright \cite{wright} who has not proposed  any additional explanations beyond a
mere fit discussion. The controversy stems from the number of data 
points used in
measuring the shortest time intervals for a recession. In order to 
justify which
law(s) better describe(s) the frequency distribution of recession periods of a
given duration a different set of data is hereby investigated. The present idea
is to consider results using a "high frequency data set", i.e. considering
quarterly, rather than annual periods as in \cite{ormer,wright}. According to
standard scaling range theories \cite{stanleybook,Bak}  the Ormerod and
Mounfield`s hypothesis, if valid, should be observable also on different time
scales.

In Sect. \ref{data} the data source is described. The data analysis for
recessions is found in Sect. \ref{recession} and that for prosperity 
durations in
Sect. \ref{prosperity}. Conclusions are found in Sect. 
\ref{conclusion}. Another
type of two parameter fit is attempted, i.e. a double exponential in 
an Appendix.

\section{Data source} 
\label{data}

A recession, for a given country, has occurred when its GDP has 
decreased between
ends of two consecutive time intervals. The recession duration may last several
time intervals (a ''period''). This definition is equivalent to that used by
Ormerod and Mounfield \cite{ormer}. Some economists might prefer
that periods with actual growth above
a time-averaged growth rate $G$
  per year or per quarter should be counted as prosperity,
and those with growth below $G$ as recession.  This surrogate data
might lead to other conclusions, but no such investigation has been 
made at this time.

The complementary set of data points is the so
called set of {\it prosperity} occurrences. In both cases no specification is
hereby made on the "strength" of the recession or prosperity. In 
other words the
time series is considered to be a set based on two characters (or + 
and - signs)
similar in physics to a magnetic (up or down) spin chain or in informatics to a
series of (0, 1) bits. In physics words, we consider the total number 
of spins in
the chain, the number of spin domains and their size (the number of 
domain walls
is the number of recession occurrences). The analysis of their distribution has
been performed in the same spirit as in \cite{ormer}. However for the sake of
comparison, the different data sets are normalised such that it is 
the frequency
$ f $ of a recession occurrence time interval which is examined rather than the
number of occurrences.

Ormerod and Mounfield \cite{ormer} and Wright \cite{wright} have 
analysed the GDP
{\it annual} records of 17 countries: Australia, Austria, Belgium, Canada,
Denmark, Finland, France, Germany, Italy, Japan, Netherlands, Norway, New
Zealand, Sweden, Switzerland, United Kingdom and USA, i.e. a total of 1965 data
points, over 124 years, i.e. between 1870 and 1994. Ormerod and Mounfield
\cite{ormer} found a power law dependence for the 336 occurrence of recessions,
lasting {\it in toto} 541 years (y)  (Tables 1 and 2). Wright \cite{wright}
reached a quite different conclusion.

In order to verify the findings it is useful to examine a different set of data
within  different time limits. Therefore GDP data for { \it quarter } period
reports are hereby examined. The data is taken from the Organisation 
for Economic
Co-operation and Development (OECD) web page \cite{oecd}, where quarterly GDP
data are found over 14 years from 1989 Q1 to 2003 Q2 (N.B. Q1, Q2, 
Q3, Q4 denotes
first, second, third and fourth quarter of a year respectively.) for 21
countries: Australia, Austria, Belgium, Canada, Denmark, Finland, France,
Germany, Italy, Japan, Korea, Mexico, Netherlands, New Zealand, 
Norway, Portugal,
Spain, Sweden, Switzerland, Great Britain and USA, amounting to a total of 1100
data points, distributed over 213 quarters (Q) for 136 recession occurrences
(Tables 1 and 2). (N.B. In the case of Germany the data are only available from
1991 Q1, Portugal from 1995 Q1 and Sweden from 1993 Q1.)

Within the paper the following notation will be used: "y" for year and "Q" for
quarter; GDP17 is  the data studied in \cite{ormer, wright}, the so called low
frequency data, while GDP21 is the so called high frequency data, used for the
present study.

The statistical properties of the low frequency $ (y^{-1}) $ and high frequency
$(Q^{-1}) $ recession duration distributions are given in  Table \ref{stat}. At
this time there are more data points in the former set, since the $Q$ recording
is more recent. However the statistical properties look quite similar. The
recession duration distribution mean, variance, kurtosis, skewness and entropy
have the same order of magnitude. Recall (Table \ref{stat}) that the 
statistical
entropy is defined as: \be S=-\sum f_i \ln(f_i). \ee

\section{Recession analysis} 
\label{recession}

Notice first that a statistical analysis and tests performed on shuffled data,
i.e. in the case of an equivalent in size set of purely stochastically
independent data taken from a skewed binomial distribution taking into account
the empirically found relative probability of recession or prosperity, implies
that such a distribution is described by an exponential law. This can be
understood since in the case of independent events the probability of 
registering
the sequence of length $ n $ of identical events is $ p(n) = P^n $, where $ P $
is the probability of the occurring event.

\subsection{Low frequency GDP17 data}

\begin{figure}[ht] 
\begin{center}
\includegraphics[scale=0.35,angle=-90]{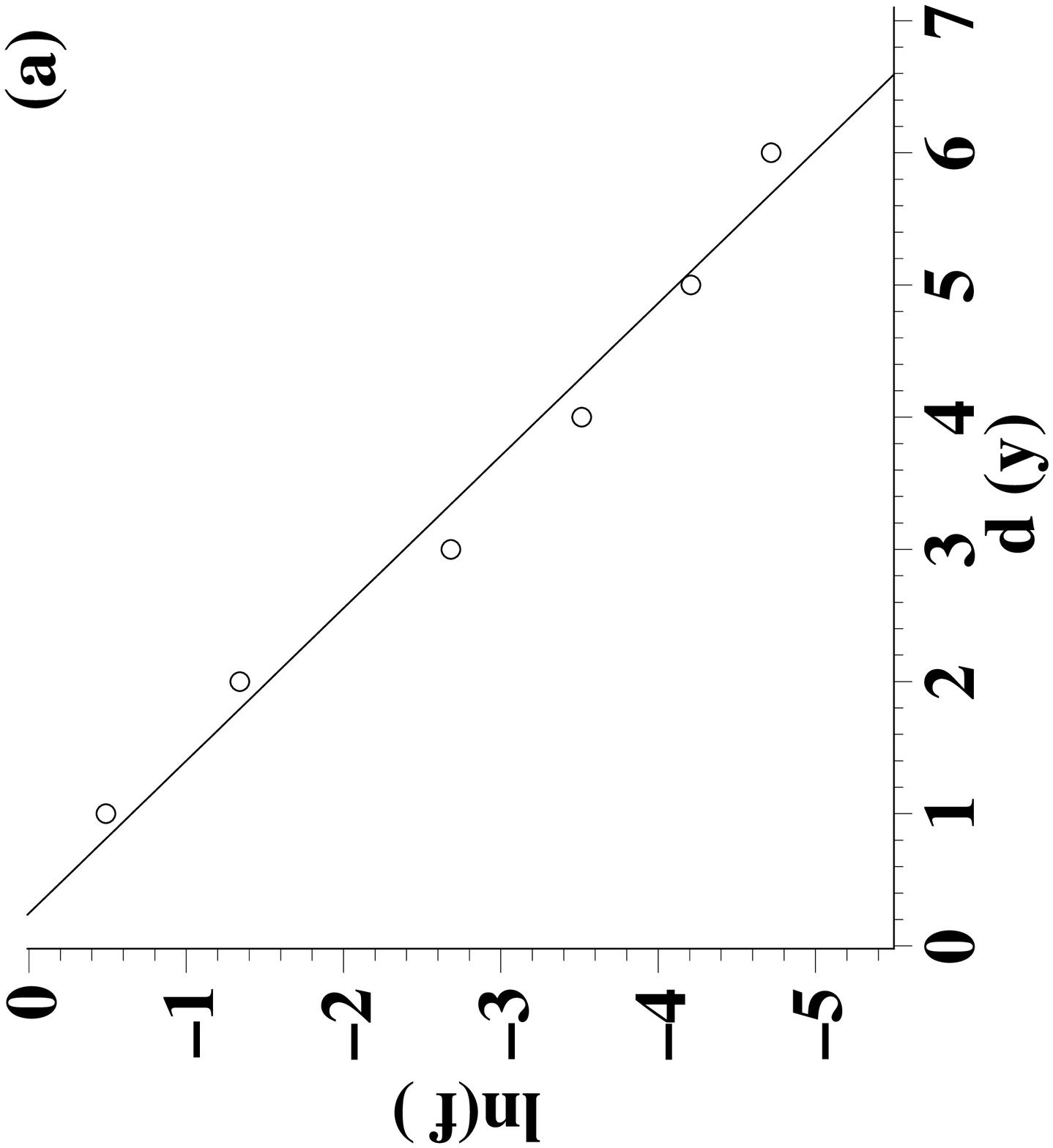}
\includegraphics[scale=0.35,angle=-90]{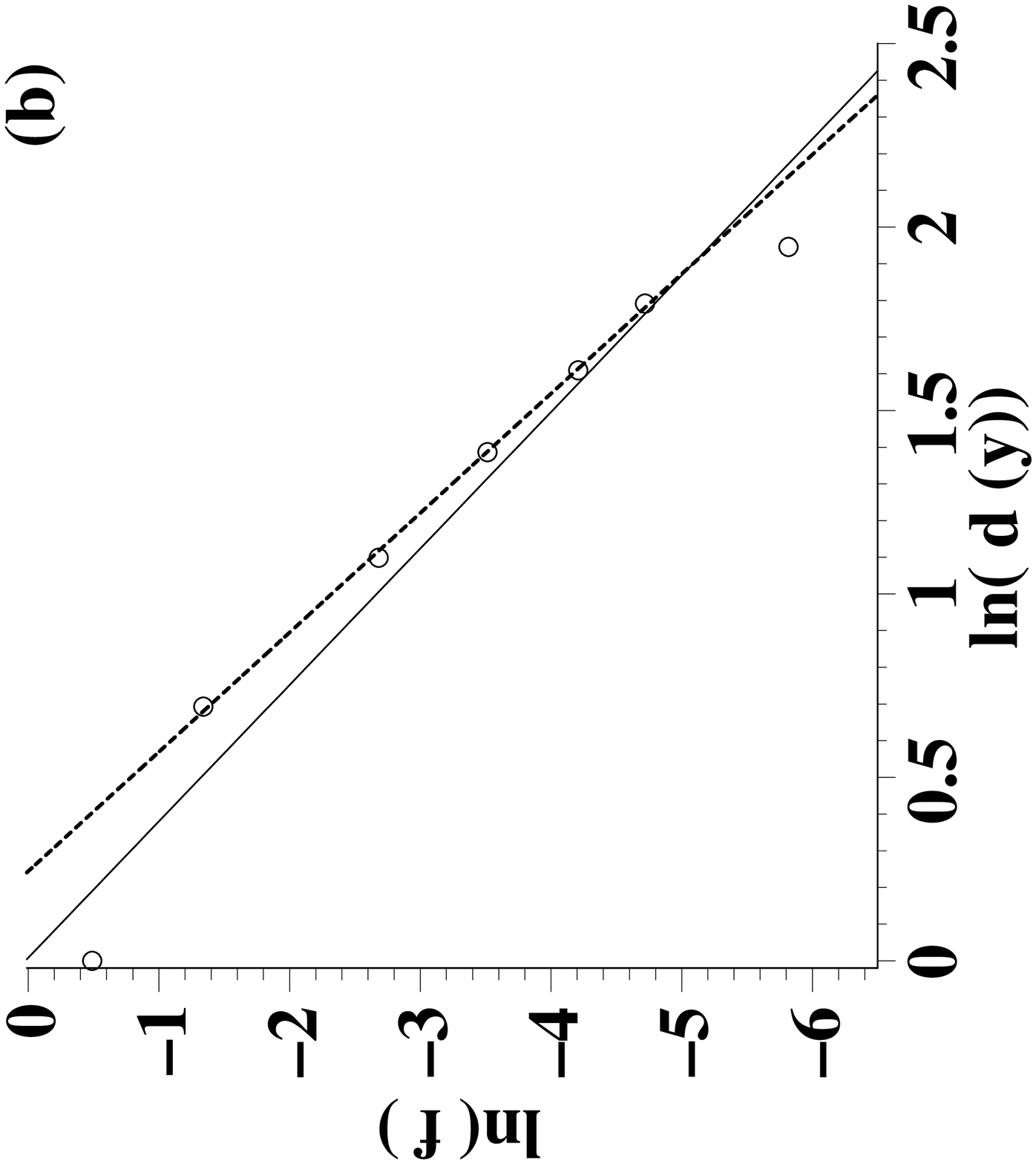} 
\end{center}
\caption{\label{art1} Data examined in \cite{ormer,wright} in (a) semi-log, (b)
loglog plot. Solid line, the best fit to all data; dashed line, the best fit to
the data without the longest and the shortest durations included, as in
\cite{wright} } 
\end{figure}

The data collected by Ormerod and Mounfield \cite{ormer} as recalled in Table
\ref{tabormer} is presented on a log-log and a semi-log plot in 
Fig.\ref{art1}(a)
and Fig.\ref{art1}(b) respectively with the best fitting straight line in both
cases, corresponding to

\be 
\label{power} 
f(d)=\gamma d^{-\delta} 
\ee 
and 
\be 
\label{exp} 
f(d)=\alpha e^{-\beta d}. 
\ee 
The power law in Eq.(\ref{power}) can be 
alternatively written as 
\be 
\label{alter} 
f(d)=(\hat{\gamma} d)^{-\delta}. 
\ee

The correlation coefficients have been calculated for both the semi-log and
log-log transformations; see Table \ref{fit} for their value and those of the
theoretical formula  parameters.

\begin{table}[ht] 
\begin{tabular}{|l|c|c|c|} 
\hline & GDP 17 &  GDP 21 & GDP 21
\\ 
& recession & recession & prosperity\\ 
& (time=y) & (time=Q)& (time=Q) \\
\hline Data length (time) & 541 & 213 & 887 \\ \hline Number of 
occurrences & 336 & 136 & 144 \\ 
\hline Mean (time) & 1.610 & 1.566 &  6.160 \\ \hline 
Variance ($time^{2} $) & 1.003 & 1.092 & 50.876 \\ 
\hline 
Kurtosis ($ time^{3} $) & 2.268 & 2.346 & 2.766 \\ 
\hline Skewness ($ time^{4} $) & 9.114 & 9.328 & 12.131\\ \hline
Entropy & $ 1.06 $ & $ 1.0 $ & $ 2.57 $  \\ 
\hline 
\end{tabular} 
\caption{ \label{stat} Statistical properties of low \cite{ormer} and high frequency
\cite{oecd} GDP data} 
\end{table}

\begin{table}[ht] 
\begin{tabular}{|l|c c c c c c c|} 
\multicolumn{8}{l}{GDP17, low frequency data (1965 data points)}   \\ 
\hline Duration (y) of recessions& 1
& 2 & 3 & 4 & 5 & 6 & 7 \\ 
\hline Number of recessions & 206 & 88 & 
23 & 10 & 5 &
3 & 1  \\ 
\hline Frequency & 0.613 & 0.262 & 0.068 & 0.030 & 0.015 & 0.009 &
0.003 \\ 
\hline \hline \multicolumn{8}{l}{GDP21, high frequency data (1100 data
points)}  \\ 
\hline Duration (Q) of recessions& 1 & 2 & 3 & 4 & 5 &6& 
7 \\ 
\hline
Number of recessions & 93 & 23 & 12 & 4 & 3 &0& 1 \\ 
\hline Frequency & 0.684 &
0.169 & 0.088 & 0.029 & 0.022 &0& 0.007  \\ 
\hline 
\end{tabular} 
\caption{
\label{tabormer} Durations of recessions. The low \cite{ormer} and 
high frequency
GDP data \cite{oecd}} 
\end{table}

\begin{table}[ht] 
\begin{tabular}{|l|c|c|c|c|} \hline & 
\multicolumn{2}{|c|}{GDP 17} &  \multicolumn{2}{c|}{GDP 21}    \\ 
& \multicolumn{2}{c|}{recession}  &
\multicolumn{2}{c|}{recession}    \\ 
& \multicolumn{2}{|c|}{ $ [ time = y ] $ }&
\multicolumn{2}{c|}{$ [ time = Q ] $}  \\ 
\hline data points & 7  & 5 
& 6 & 5 \\
\hline R ({\it semi-log}) & -0.993  & -0.983  & -0.823 & -0.858  \\ 
\hline R
({\it log-log}) & -0.976 & -0.9996 & -0.937& -0.918 \\ 
\hline $ \ln(\alpha) $ & $
0.21 \pm 0.21 $ & $ 0.02 \pm 0.38 $ & $ -0.13 \pm 0.33 $ & $ 0.20 \pm 
0.34 $   \\
\hline $ \alpha $ & $ 1.24 $ & $ 1.02 $ & $ 0.88 $ & $ 1.22 $ \\ 
\hline $ \beta [
time^{-1} ] $ & $ 0.87 \pm 0.05 $  & $  0.83 \pm 0.09 $ & $ 0.73 \pm 0.08 $ & $
0.86 \pm 0.10 $ \\ 
\hline $ \delta $ & $ 2.69 \pm 0.27 $ & $ 3.07 \pm 
0.05 $ & $
2.30 \pm 0.14 $ & $ 2.17 \pm 0.18 $  \\ 
\hline $ \ln(\gamma) $ & $ 0.02 \pm 0.37
$ & $ 0.75 \pm 0.07  $ & $ -0.22 \pm 0.19 $ & $ -0.31 \pm 0.18 $ \\ 
\hline $\gamma [time^{-\delta} ] $ & $ 1.02 $ & $ 2.11 $ & $ 0.80 $ & $ 0.73 
$ \\ 
\hline
$ \hat{\gamma} [time^{-1} ] $ & $ 0.99 $ & $ 0.78 $ & $ 1.10 $ & $ 1.16 $ \\
\hline $ 1// \zeta (\delta)  $ & $ 0.78 $ & $ 0.84  $  & $ 0.69 $ & $ 
0.66  $ \\
\hline 
\end{tabular} 
\caption{ \label{fit} Recession data; fitting parameters}
\end{table}

\begin{table}[ht] 
\begin{tabular}{|l|c|c|} \hline & \multicolumn{2}{c|}{GDP 21}
\\ 
& \multicolumn{2}{c|}{prosperity}  \\ 
\hline data points &  23 & 15  \\ 
\hline
R ({\it semi-log}) & -0.976 & -0.980 \\ 
\hline R ({\it log-log})  & -0.992 &
-0.992 \\ 
\hline $ \ln(\alpha) $ & $ -2.55 \pm 0.23 $ & $ -1.97 \pm 0.23 $ \\
\hline $ \alpha $ & $ 0.08 $ & $ 0.14 $ \\ 
\hline $ \beta [ Q^{-1} ]$ 
& $  0.079
\pm 0.012 $ & $ 0.13 \pm 0.022 $ \\ 
\hline $  \delta $ & $  1.12 \pm 0.09 $ & $
1.01 \pm 0.12 $ \\ 
\hline $ \ln(\gamma) $ & $ -1.12 \pm 0.23 $ & $ 
-1.23 \pm 0.25
$  \\ 
\hline $ \gamma [Q^{-\delta} ] $ & $ 0.33 $ & $ 0.29 $ \\ 
\hline $
\hat{\gamma} [Q^{-1} ] $ & $ 2.43 $ & $ 2.74 $ \\ 
\hline $ 1//\zeta 
(\delta) $  &
$ 0.11 $ & $ 0.01  $ \\ 
\hline 
\end{tabular} 
\caption{ \label{fitprosp}
Prosperity data; fitting parameters} 
\end{table}

In the case of data presented in \cite{ormer,wright} the best correlation
coefficient (Table \ref{fit}) occurs for $R_{GDP17,7_{semi-log}}=-0.993$ to be
contrasted to the value  $R_{GDP17,7_{log-log}}= -0.976$, supporting the
hypothesis \cite{wright} that the data follows an exponential law.

If two (here, rather than one as in \cite{wright}) data points corresponding to
the longest  and the shortest periods are dropped,  a linear fit on the log-log
plot appears significantly better (see dash line in Fig.\ref{art1}(b)). Indeed
the correlation coefficients take values $R_{GDP17,5_{semilog}}=- 0.983$, and $
R_{GDP17,5_{loglog}}=-0.9996 $ respectively, suggesting that the shortest data
follows a power law, as in \cite{ormer}.

Thus the case of short and long durations should receive better 
attention. Due to
the scarcity of data points only a shorter time scale is convenient. 
Indeed very
"long period" cases, like over decades, are not easily available nor numerous.
Moreover it might be necessary to distinguish them according to the 
depth of the
recession. In fact the (annual) data contains only one such a case (Table
\ref{tabormer}). Thus we consider only "higher frequency data" even though such
(quarterly) GDP reports are necessarily less numerous, since in fact 
they belong
to the most recent times.

\subsection{High frequency GDP21 data}

The GDP data for the occurrences of recessions and their duration is 
presented in
Table \ref{tabormer} for 21 countries \cite{oecd}. As in the case of 
data used in
\cite{ormer} the highest number (and frequency) of recessions is registered for
the case of the shortest period. The longest recession (7 Q) occurred 
in Finland;
there are several five quarter long recessions exactly corresponding 
to those in
the yearly data (Fig. 2). The statistical characteristics of the 
recession period
data are presented in Table \ref{stat}. The overall relationship between the
duration and occurrence of recessions seems to be similar in both (high and low
frequency) sets of data, though the quarterly data appears to be more peaked
(Table \ref{tabormer}).

The collected data are presented in semi-log (Fig.\ref{fig4}(a)) and log-log
(Fig.\ref{fig4}(b)) plots together with the best fitting lines. The 
parameters of
Eqs. (\ref{power})- (\ref{exp}) and the correlation coefficients are given in
Table \ref{fit} and Table \ref{fitprosp} both for the semi-log and 
log-log fits.
Comparing the values of the correlation coefficients $ R_{GDP21_{-}semilog}
=-0.823 $ and $ R_{GDP21_{-}loglog =-0.937} $ (Table \ref{fit}) we may conclude
that the data from \cite{oecd} better follows a power law as found in
\cite{ormer}.

\begin{figure}[ht] 
\begin{center}
\includegraphics[scale=0.35,angle=-90]{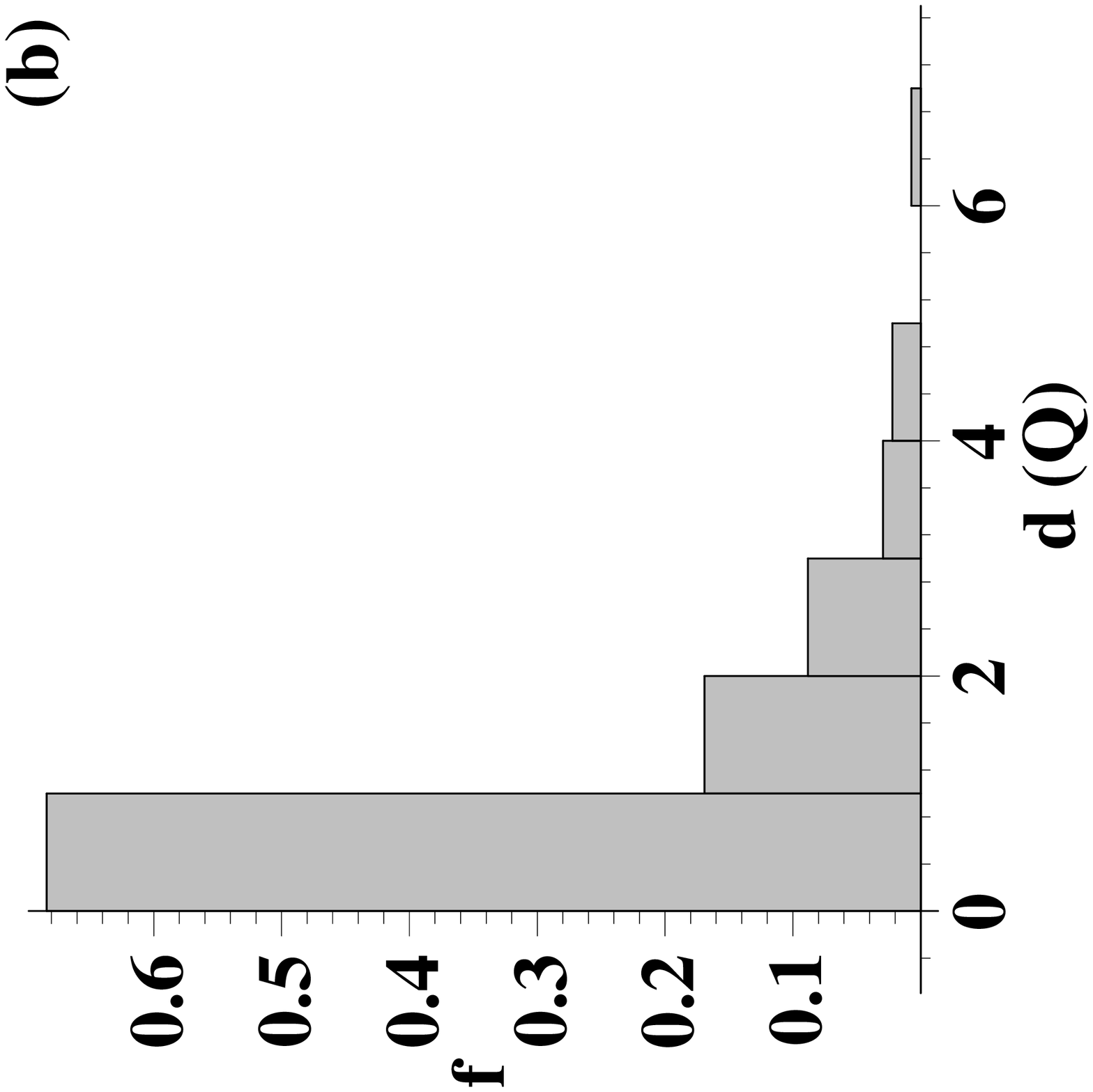}
\includegraphics[scale=0.35,angle=-90]{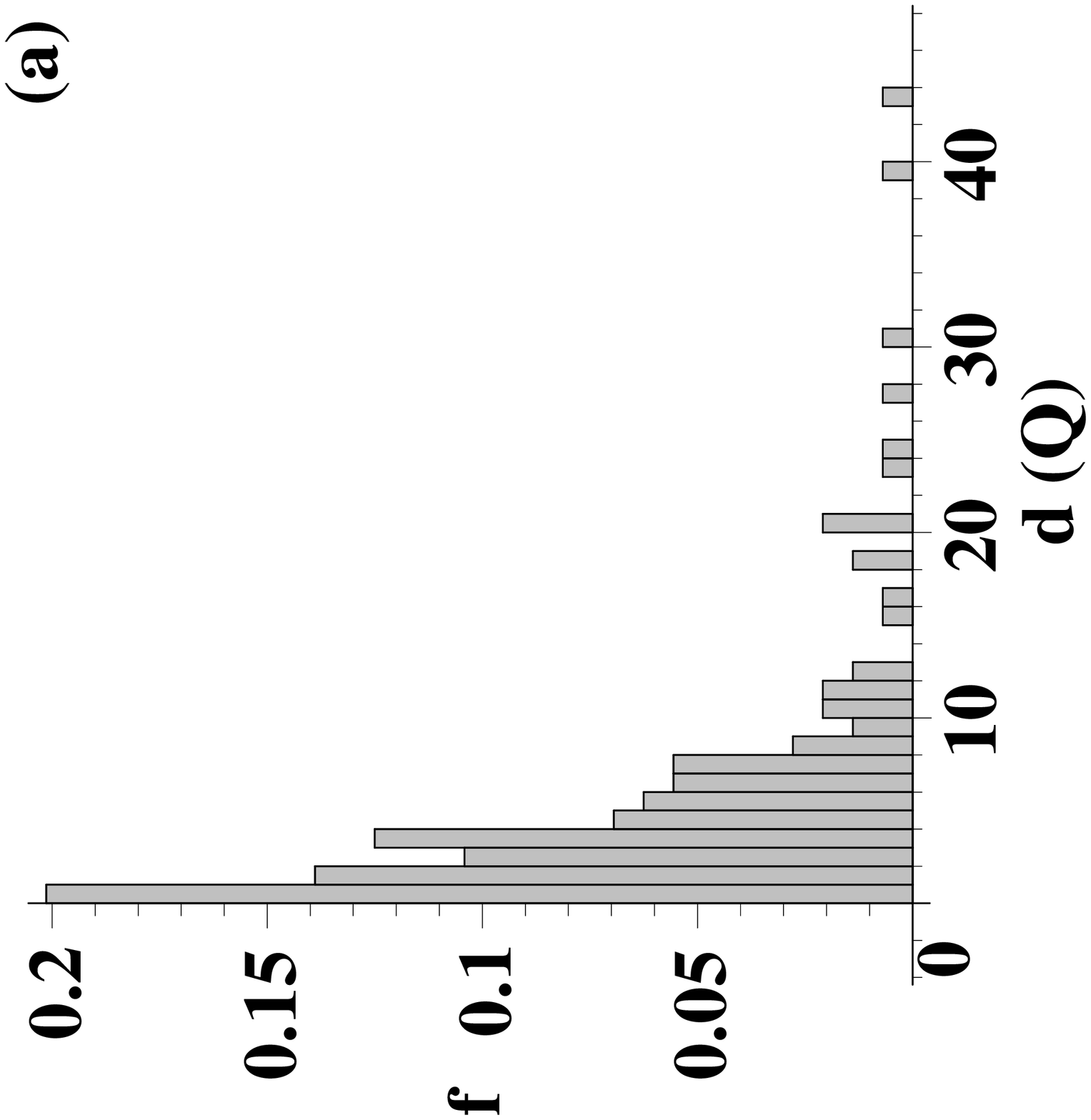} 
\end{center}
\caption{\label{hist1} (a) Histogram of recession durations observed 
in the GDP21
data \cite{oecd}; \label{fig8} (b) Histogram of the prosperity periods}
\end{figure}

\begin{figure}[ht] 
\begin{center}
\includegraphics[scale=0.35,angle=-90]{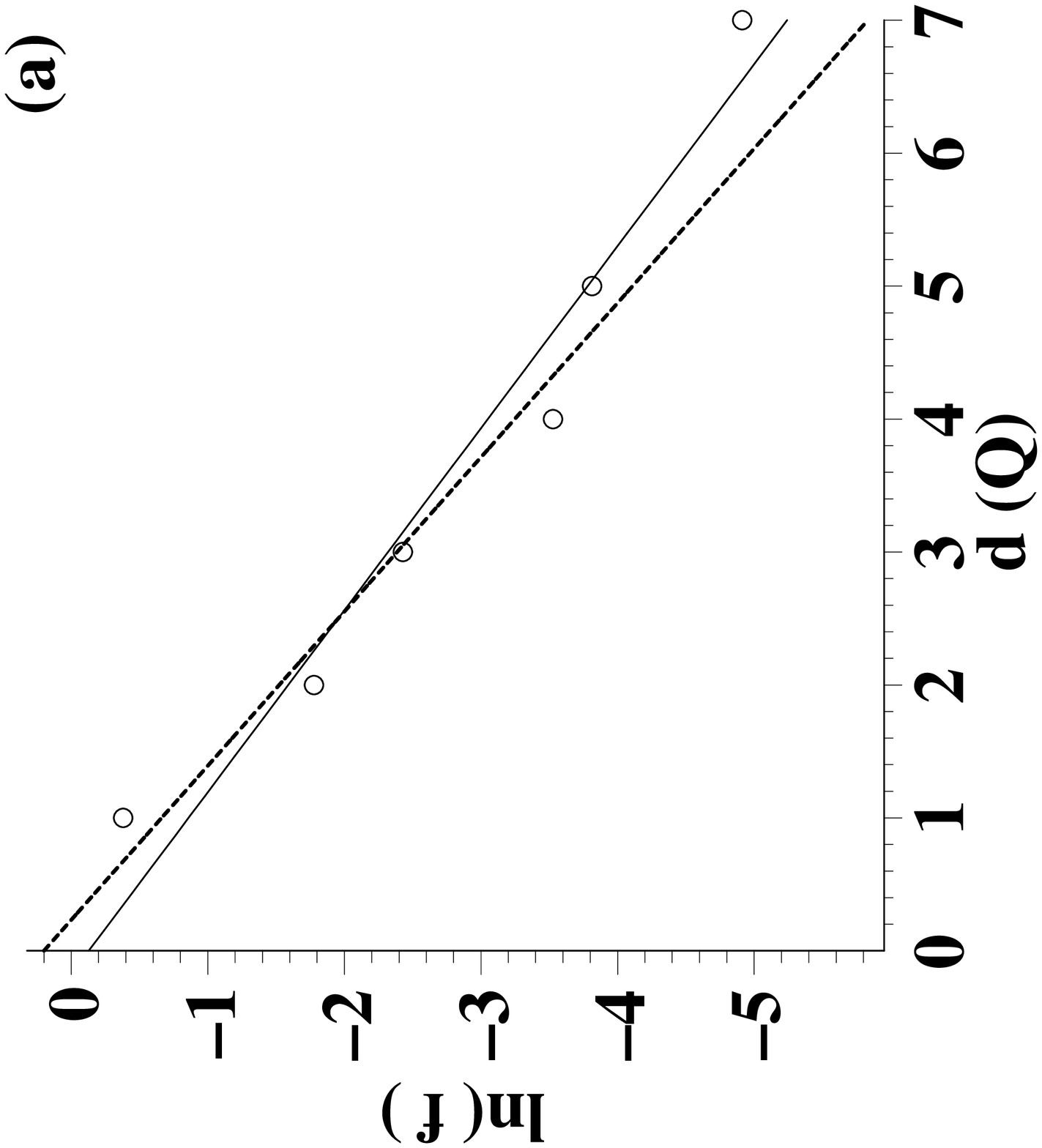}
\includegraphics[scale=0.35,angle=-90]{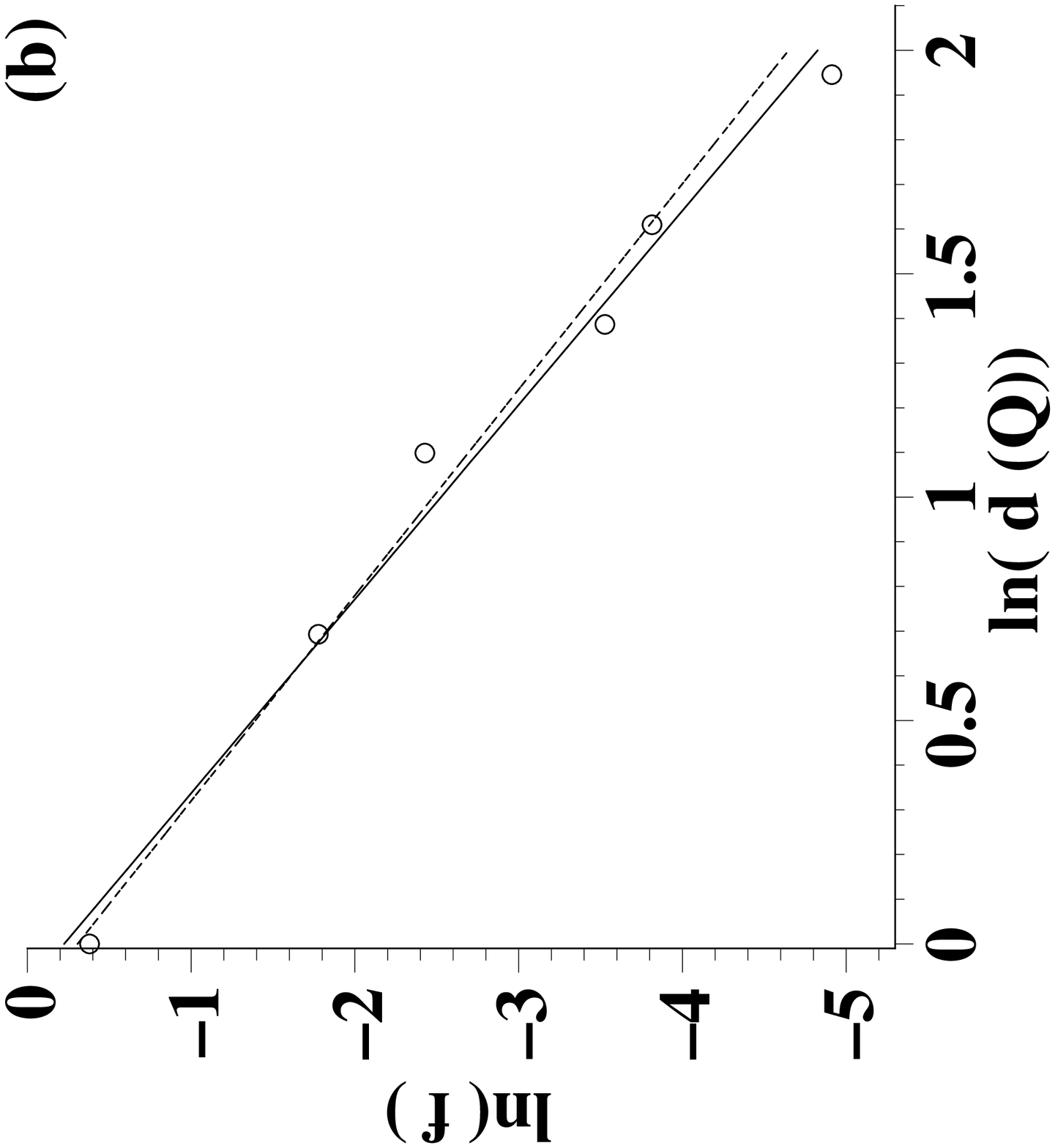} 
\end{center}
\caption{\label{fig4} Recession. Collected GDP21 data in (a) a semi-log and (b)
log-log plot together with the fitted line. Solid line - all data, dashed one -
the data without taking into account the longest recession duration} 
\end{figure}

\subsection{Truncated data}

Following the ideas of \cite{wright}, let us remove the rarest case. The fits
(Fig. \ref{fig4}) are not changing significantly; the correlation coefficients
(Table \ref{fit}) do not much improve : $ 0.82 \rightarrow 0.86 $, $ 0.94
\rightarrow 0.92 $. However, since $|R_{GDP21,5_{-}log-log} |<
|R_{GDP21,6_{-}log-log}| $, ($0.918<0.937$) in this case,  a strong 
argument  for
the power law seems to hold for the  GDP21 data. The parameters of 
the laws (Eq.
(\ref{power}) - (\ref{exp})) are given in Table \ref{fit} as well.

\section{Prosperity analysis} 
\label{prosperity}

Investigating the problem of dependencies between the frequency of 
occurrence and
the duration of recessions another question arises: is there any law governing
prosperity duration times?

A period is treated as a prosperity one if the GDP has increased 
between  the end
of two consecutive time intervals. The prosperity periods are complementary to
the recession periods. There are 887 positive Q's (or ''up-spins'') distributed
into 144 periods (''domains''), Table \ref{stat}. E.g. the prosperity duration
may be much longer than recession durations, as up to 44 Q (for Great Britain).
The occurrence of durations of prosperity periods is presented as a 
histogram in
Fig. \ref{fig8} while the statistical properties  are given in Table 
\ref{stat}.
The mean value of the prosperity periods is about four times longer than in the
case of the recession durations, for the time intervals examined here.

\begin{figure}[ht] 
\begin{center}
\includegraphics[scale=0.35,angle=-90]{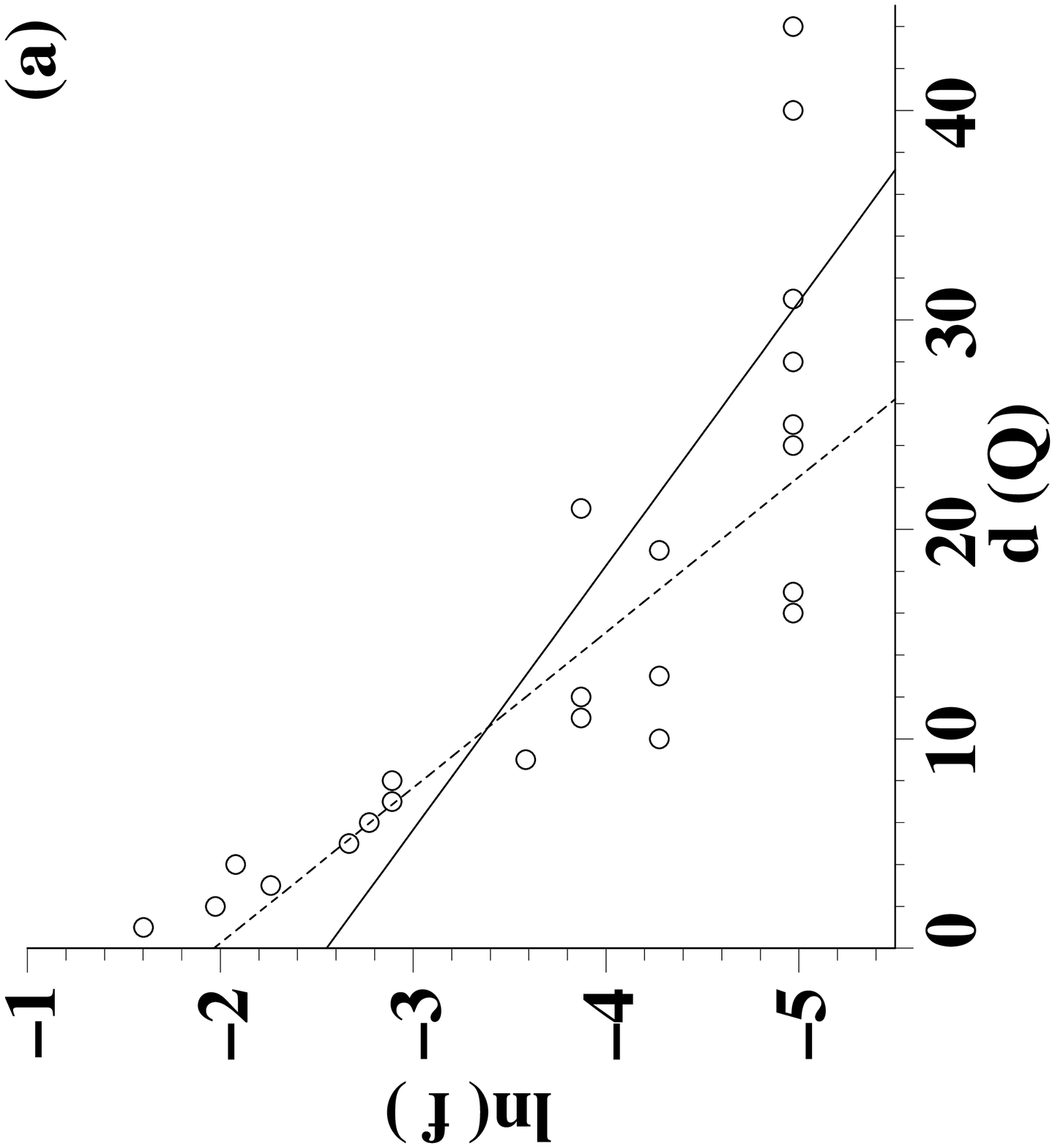}
\includegraphics[scale=0.35,angle=-90]{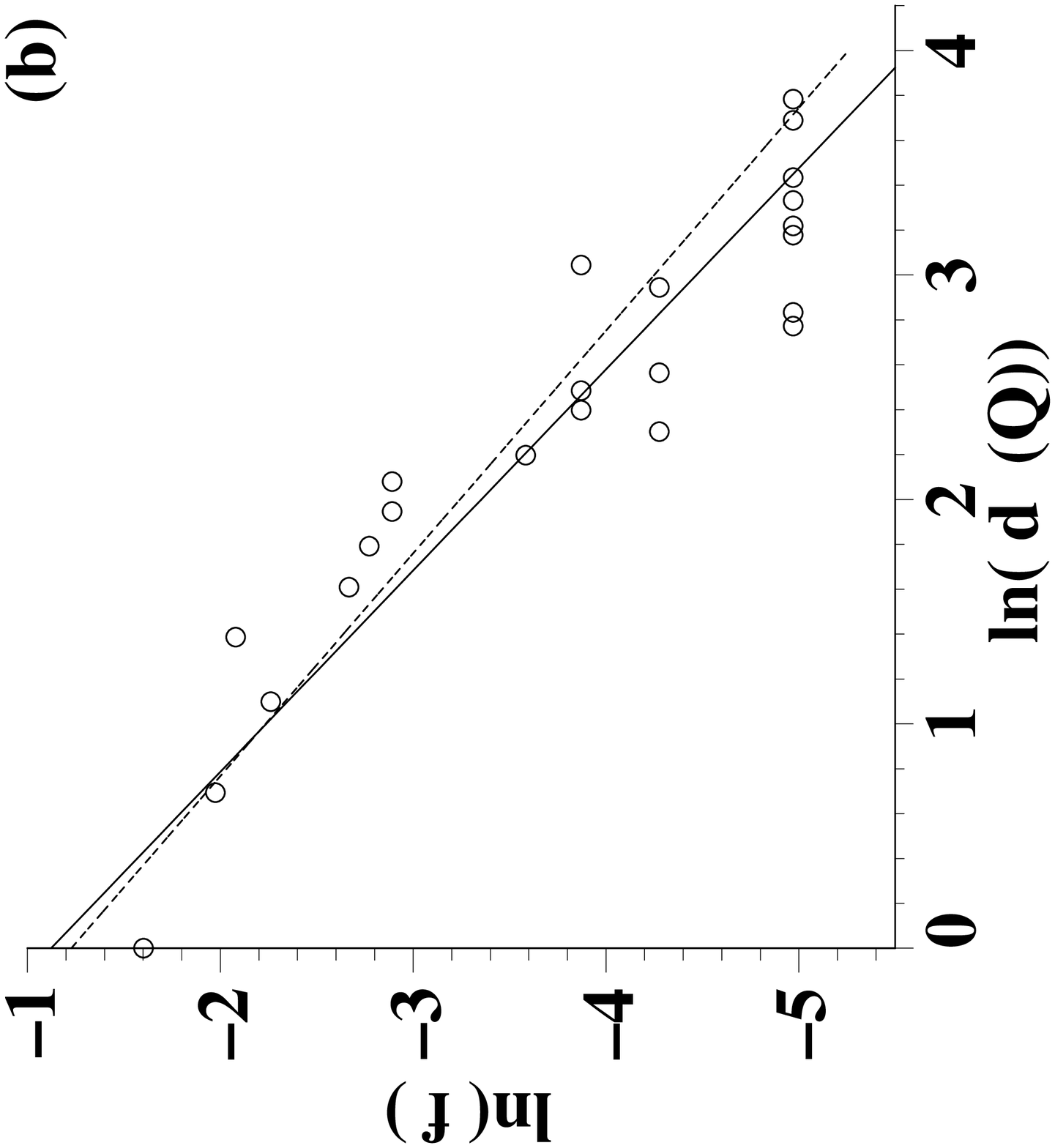} 
\end{center}
\caption{\label{fig5} Collected GDP21 data of prosperity duration 
occurrences in
(a) semi-log and (b) log-log plot together with the fitted line. Solid - all
data, dashed one - the data without the cases where only one events was
registered } 
\end{figure}

The data is presented in semi-log (Fig.\ref{fig5}(b)) and log-log
(Fig.\ref{fig5}(a)) plots with the best fitting lines. As in the case of the
recession time distribution, the correlation coefficients have been calculated
together with the values of the parameters appearing in the functions
Eq.(\ref{exp}) and Eq.(\ref{power}). The results are found in Table
\ref{fitprosp}.

The investigated data contains eight cases (durations (Q) = 16; 17; 24; 25; 28;
31; 40; 44) where only one occurrence of such a size is registered. The fit has
been repeated on the corresponding truncated data set, in line with 
\cite{wright}
idea. The results are found in Table \ref{fitprosp}. The fit precision does not
(to say the least) increase, and a power law still better describes 
the data. The
parameters and regression coefficients of the power law Eq.(\ref{power}) are
given in Table \ref{fitprosp}) as well.

\begin{figure}[ht] 
\begin{center}
\includegraphics[scale=0.5,angle=-90]{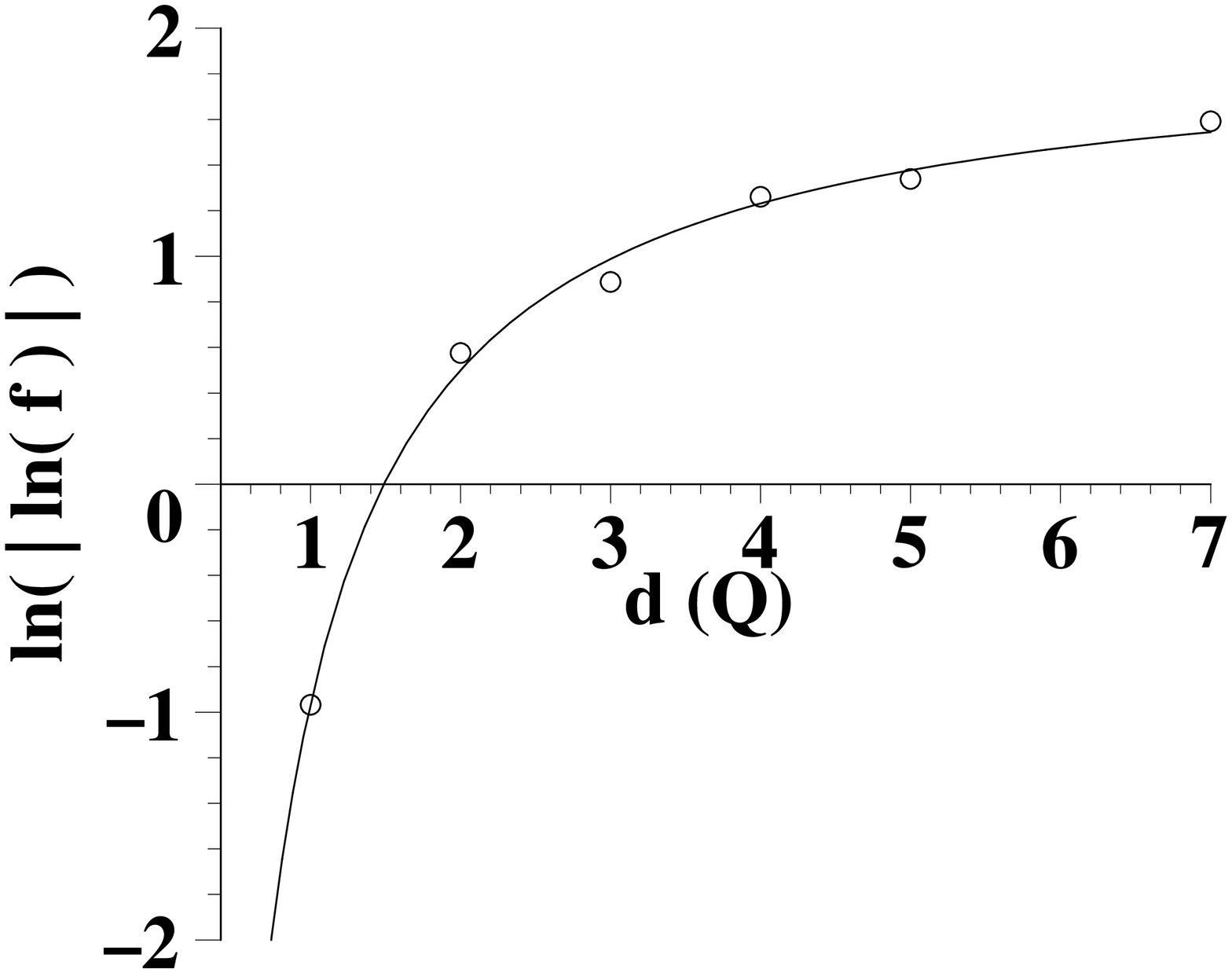} 
\end{center}
\caption{\label{fig6} Collected GDP21 data of recession duration frequency
occurrences in normal-loglog plot together with the fitted curve, $ |ln(f)|
=\lambda e^{-\frac{\mu}{d}} $} 
\end{figure}

\section{Conclusions} 
\label{conclusion}

The observed correlations between the duration of a recession (or prosperity)
periods and their probability of occurrence is hereby interesting. It appears
that the relationship is nontrivial.  First the values of the exponents
$\beta^{-1}$ and $\hat{\gamma}^{-1}$  give an information on the ''relaxation
time'' of the processes. It is roughly 1.0 (y or Q)  for {\it 
recessions}; this
is a remarkable scale free, whence so called universal, result. However the
corresponding "relaxation time" is an order of magnitude different, 
i.e. 10 Q for
{\it prosperity} cases. There is a major difference concerning  the {\it decay
exponent} $\delta$ though: $\delta$ is about -2.5 for recessions, but 
$\sim$ -1.1
for the prosperity cases. Also, $\hat{\gamma}^{-1}$ is quite different from
$\beta^{-1}$ for the prosperity cases. This points again to non universal
features (or to different universality classes ?).

From an economic point of view the above relaxation time values should not be
confused with the apparent periodicity (or better ''periodicities'') 
of business
cycles 
\cite{Kalecki1,Kalecki2,Kalecki3,others1,others2a,others2b,others2c}. 
The
latter ones are rather to be compared to the means found in Table 1. Even these
seem small with respect to  so called ''common feeling'', which 
rather measures a
median more than a mean. In some sense this indicates the difference between a
psychological or even visual data filtering based on the slowliest trend,
mathematically a large window-moving average, in contrast to the actual works
looking at somewhat higher frequency data.

Along this psychological line of expectations for economy periods,
Hohnisch et al. \cite{Hohnisch} discuss the ups and downs
of $opinions$ about recessions and prosperity along a Blume-Capel model and
Glauber dynamics. Their computer simulations  show that opinions
drastically undergo changes from one equilibrium to the other,
both having  rough but small
fluctuations, - as in stochastically resonant systems.

Nevertheless, except for the debatable GDP17 full data set, considering the
results obtained in the above sections \ref{recession} and 
\ref{prosperity} it is
observed that a power law significantly better describes the distribution of
durations of both recession and prosperity periods. For recessions, 
another type
of two parameter fit can be also attempted, i.e. a double exponential, see
Appendix.

A final argument in favor of a true power law distribution should 
follow from the
measure of the  $\gamma$ parameter which should be equal to $1//\zeta(\delta)$
where $ \zeta (\delta)$ is the Riemann $\zeta$ function. The $1//\zeta(\delta)$
values are given as the last line in each Table. It is remarkable in 
view of the
available data  and the error bars that there is some reasonable agreement for
the recession cases, the more so for the high frequency data. The prosperity
cases do not seem to obey such a criterion. This is of course due to the huge
tail of rare (though very long) prosperity events during the last years.

In summary the hypothesis stated by Ormerod and Mounfield seems to be fine for
recession occurrences though a theory is still needed, but the case 
of prosperity
durations is not settled, due to the very long and low tail of the 
distribution.

The power law hereby preferred thus shows some coupling
between successive (economy or opinion or..) swings is
missing in the (simple) computer model of Hohnisch et al. 
\cite{Hohnisch}, like the (simple)
random walk model of Bachelier \cite{Bachelier1900} which gives a 
Gaussian or log-normal distribution
of the price fluctuations, needs
a non-random explanation, like psychological herding etc. \cite{stauffer}
for  getting as  actually here the observed power law distribution.
It is finally fair to mention that such successions of recessions and 
prosperity
periods, sometimes called business cycles, have recently received a 
renewed deal
of attention from analytical and simulation points of view
\cite{others2a,others2b,others2c,ACP1,ACP2,ACP3,JMMA1,JMMA2} beside theoretical
work in classical macroeconommic studies \cite{macro}.

\vskip 1cm

{\bf Appendix}

Another type of two parameter fit can be also attempted, i.e. a double
exponential ($ee$): \be \label{doublexp} |ln(f)| =\lambda 
e^{-\frac{\mu}{d}}, \ee
where $ f $ denotes the occurrence frequency and $ d $ the duration of a
(recession or) prosperity period, while $ \mu $, and $\lambda $ are unknown
coefficients. In the case of GDP21 recession data:  $ \mu =  2.92 $ Q, and $
\lambda= 7.11 $ while the correlation coefficient is $ R_{ee_{-}loglog} =
-0.998$, - the best of all attempted fits when the full data is used 
in the main
text. The result seems to be interesting, but there is no theoretical
interpretation of it. However notice that the formula looks like that
corresponding to an Arrhenius formula in chemistry where the $\mu$ coefficient
corresponds to an activation energy, when $d$ is the temperature.

This (Arrhenius-like) formula usually applies when some system 
balances between a
stable state and an excited state. In that spirit one might consider that the
value of $\mu \simeq 3 Q$ is of the same order of magnitude as the 
value   (1 Q)
found in the main text for the power law, and is in the range estimated as a
realistic time necessary for responding to some $(de)stabilisation$ 
constraint of
the economic field. The double exponential form in financial data has received
some consideration in \cite{takayasudoublexp}.

\section{Acknowledgements} This work is partially financially supported by FNRS
convention FRFC 2.4590.01. J. M. would like also to thank GRASP for the welcome
and hospitality. Comments by an anonymous referee who
pointed out ref. \cite{Hohnisch}
are greatfully acknowledged.

\end{document}